\documentclass[aps,prd,twocolumn,nofootinbib]{revtex4}
\usepackage{graphicx}
\usepackage{hhline}
\usepackage[mathscr]{euscript}
\usepackage{outlines}
\usepackage{xcolor}
\usepackage{placeins}
\usepackage{amssymb, amsmath}

\newcommand{\Ms}{M_{\odot}}

\usepackage{tikz}
\usepackage{pgfplots}
\usepackage{hyperref}

\newcommand{\mnras}{Mon.\ Not.\ Roy.\ Astron.\ Soc.}

\newcommand{\aap}{Astron.\ and Astrophys.}

\newcommand{\araa}{Annual Reviews of Astr.\ and Astrophys.}

\usepackage[normalem]{ulem}

\newcommand{\coderefs}{\cite{Duez:2005sg,Paschalidis2011,Etienne2012,Etienne2015}}

\begin{document}

\title{Improving the convergence order of binary neutron star merger simulations in the BSSN formulation}

\author{Carolyn A. Raithel,$^{1,2,3}$ Vasileios Paschalidis,$^{4,5}$}
\affiliation{$^1$School of Natural Sciences, Institute for Advanced Study, 1 Einstein Drive, Princeton, NJ 08540, USA}
\affiliation{$^2$Princeton Center for Theoretical Science, Jadwin Hall, Princeton University, Princeton, NJ 08540, USA}
\affiliation{$^3$Princeton Gravity Initiative, Jadwin Hall, Princeton University, Princeton, NJ 08540, USA}
\affiliation{$^4$Department of Astronomy and Steward Observatory, University of Arizona, 933 N. Cherry Avenue, Tucson, Arizona 85721, USA}
\affiliation{$^5$Department of Physics, University of Arizona, 1118 E. Fourth Street, Arizona 85721, USA}

\begin{abstract}

High-accuracy numerical relativity simulations of binary neutron star
mergers are a necessary ingredient for constructing gravitational
waveform templates to analyze and interpret observations of compact
object mergers. Numerical convergence in the post-merger phase of such
simulations is challenging to achieve with many modern codes. In this
paper, we study two ways of improving the convergence properties of
binary neutron star merger simulations within the
Baumgarte-Shapiro-Shibata-Nakamura formulation of Einstein's
equations. We show that discontinuities in a particular constraint
damping scheme in this formulation can destroy the post-merger
convergence of the simulation. A continuous prescription, in contrast,
ensures convergence until late times. We additionally study the impact
of the equation of state parametrization on the pre- and post-merger
convergence properties of the simulations. In particular, we compare
results for a piecewise polytropic parametrization, which is commonly
used in merger simulations but suffers unphysical discontinuities in
the sound speed, with results using a ``generalized" piecewise
polytropic parametrization, which was designed to ensure both
continuity and differentiability of the equation of state. We report
on the differences in the gravitational waves and any spurious
pre-merger heating, depending on which equation of state
parametrization is used.
\end{abstract}

\maketitle

\section{Introduction}

With the advent of gravitational wave astronomy, binary neutron star
mergers offer an exciting new laboratory for studying a wide range of
topics, from the dense matter equation of state to the production of
heavy elements via the r-process, and can even be used as a standard
siren for cosmological measurements. Although each of these pursuits
is sensitive to different information contained within an observed
merger, they all rely (at least in part) on precise modeling of the
gravitational waveform to determine the source properties of the
binary.

Constructing robust gravitational waveforms for infering source
properties is a challenging problem. While analytic and semi-analytic
waveform templates suffice during the early inspiral of a neutron star
coalescence \cite{Hinderer2008,Binnington2009,Damour2009,Damour2014},
non-linear hydrodynamical effects become significant in the final
orbits.  As a result, from the late inspiral onwards, analytic
waveforms must be calibrated with fully non-linear numerical
relativity calculations, which solve the Einstein equations coupled to
the equations of (magneto-)hydrodynamics and neutrinos (e.g.,
\cite{Dietrich2017,Kawaguchi2018,Nagar2018,Dietrich2019a}).  For
recent reviews on the status of numerical relativity simulations of
binary neutron star mergers, see, e.g.,
\cite{2017CQGra..34h4002P,Baiotti2017,Duez2019,Radice2020,Ciolfi:2020cpf}.

During the inspiral, numerical relativity simulations generally show
rigorous self-convergence, typically with errors converging to zero at
second- or third-order with resolution, depending on the details of the
numerical scheme (e.g.,
\cite{Radice2014,Radice2014a,Bernuzzi2016,East2016a,Kiuchi2017,Most2019}).
However, numerical convergence is challenging to establish in the
post-merger regime (but see \cite{Most2019}). The challenge is in part
due to the turbulent nature of the post-merger evolution, but may also 
be related to aspects of the spacetime evolution, as has been found
for binary black holes~\cite{Zlochower:2012fk,Etienne:2014tia}.

In this paper, we investigate two ways of improving numerical
convergence in the post-merger phase of a binary neutron star merger,
within the Baumgarte-Shapiro-Shibata-Nakamura (BSSN) formulation of
Einstein's equations.  First, we study how the treatment of parabolic
damping of the Hamiltonian constraint can affect the global numerical
convergence of the Hamiltonian constraint and of the gravitational wave
signals. One might expect that because the Hamiltonian constraint
must be zero in the continuum limit, it might not matter how the
damping of this constraint is implemented. However, 
the challenge of treating parabolic damping arises
specifically for codes with adaptive-mesh-refinement grids, in which
adjacent refinement levels may have different degrees of numerical
dissipation due to differences in the grid spacing. As we will describe in
further detail below, the jumps in effective damping at the refinement
level boundaries introduce discontinuities in the associated evolution
equations, which can reduce the convergence order of global
quantities.

In a separate context, it was recently reported that replacing 
the standard, discontinuous
Kreiss-Oliger dissipation prescription with a continuous prescription
was necessary to achieve long-term stability in numerical simulations
of charged binary black holes \cite{Bozzola2021}. In this paper, we
will demonstrate that discontinuities in a parabolic damping of the
Hamiltonian constraint across different refinement levels can also
spoil the convergence of binary neutron star merger simulations in the
post-merger phase, using the dynamical spacetime and
(magneto-)hydrodynamics code
of Refs.~\cite{Duez:2005sg,Paschalidis2011,Etienne2012,Etienne2015}, as it
was most recently extended in~\cite{Raithel2021}. We show that by
adopting a continuous prescription for the parabolic term that is used
to damp Hamiltonian constraint violations in this framework, we
recover convergent behavior in several quantities until late times post-merger.

We stress that the results found in this paper are specific to the
BSSN formulation, but we expect them to hold also for more general
types of parabolic constraint damping at the analytic level of
formulations of the Einstein
equations
\cite{Paschalidis:2007ey,Paschalidis:2007cp}.  On the other
hand, improved constraint damping may also be achieved with different
formulations which implement lower-order constraint damping terms,
such as the conformal Z4 
\cite{Bernuzzi2010,Hilditch2013,Alic2013} 
or
the generalized harmonic
\cite{Gundlach2005,Pretorius2005,Lindblom2006}
families of
formulations.
   
As a second main goal of this paper, we investigate how the smoothness
of the dense-matter equation of state (EoS) affects the convergence
order of these simulations. In particular, we perform simulations
using the same EoS model, parameterized with either piecewise
polytropes (PWP) \cite{Ozel2009,Read2009} or with the
recently-proposed generalized piecewise polytrope parametrization
\cite{OBoyle2020}. Although piecewise polytropes are commonly used in
the merger literature, they are discontinuous in the first derivative
of the pressure and hence in the sound speed. These discontinuities
are purely artifacts of the construction, and are of potential concern
for two reasons. First, as discussed in
\cite{Leveque2002,Voss2005,Paschalidis2011}, when the fluxes of
hydrodynamic variables are non-smooth, the equations of hydrodynamics
are non-convex and this can lead to unphysical solutions such as split
waves and composite structures. Additionally, in the context of
spectral methods, Ref.~\cite{Foucart2019} recently showed that using a
smooth EoS can significantly improve the accuracy of the inspiral
waveform, compared to simulations that use a piecewise polytropic EoS.
In that work, the authors used a spectral expansion of the EoS to
ensure smoothness in the pressure.  The generalized PWP construction,
which we explore in this work, in a finite-difference/finite-volume
numerical approach, likewise ensures that the sound speed remains
continuous at all densities \cite{OBoyle2020}.

In this paper, we perform the first numerical simulations of binary
neutron star mergers with the generalized PWP construction and we show
robust self-convergence in the inspiral gravitational waves using this
parameterization, as well as convergence of the Hamiltonian constraint
until late times post-merger. We find similar gravitational wave phase
errors during the inspiral with both the standard and smoothed PWP
parametrizations. In contrast, we find a difference of $\sim130$~Hz in
the post-merger gravitational wave spectra between these
parametrizations, which cannot easily be explained by differences in the
radii or tidal deformabilities of the neutron stars. This could point
to a subtle sensitivity of the post-merger gravitational waves on the
smoothness of the underlying EoS that we plan to explore further in
forthcoming work.  

The outline of the paper is as follows. In Sec.~\ref{sec:methods}, we
describe our numerical methods, starting with a brief overview of the
evolution equations in the BSSN formulation. In
Sec.~\ref{sec:damping}, we describe three different prescriptions
for damping the Hamiltonian constraint, while in Sec.~\ref{sec:eos} we
describe the two EoS parametrizations that we explore in this work. We
present our findings in Sec.~\ref{sec:results}, starting in
Sec.~\ref{sec:conserved} with a comparison of global quantities for
each of the Hamiltonian constraint damping treatments and EoS
parametrizations.  In Sec.~\ref{sec:conv}, we discuss the convergence
of the Hamiltonian constraint for each study, and in
Sec.~\ref{sec:gw}, we analyze the gravitational wave signals from all
of our simulations. Appendix~\ref{sec:EOSfit} provides additional
details about the smoothed EoS parametrization.

Unless otherwise stated, we use geometrized units in which $G=c=1$.

\section{Numerical methods}
\label{sec:methods}
All simulations in this paper were performed with the dynamical
spacetime, general-relativistic (magneto)-hydrodynamics code with
adaptive mesh refinement
of~\cite{Duez:2005sg,Paschalidis2011,Etienne2012,Etienne2015} as it
was most recently extended in~\cite{Raithel2021}. The code is built
within the Cactus/Carpet framework
\citep{Allen2001,Schnetter2004,Schnetter2006}. Here, we review only a
few key aspects of the code, in order to highlight the changes
introduced in this paper.

In the code, the spacetime is evolved using the BSSN formulation of
the Einstein equations \citep{Shibata1995,Baumgarte1999}, in which the
evolution equations are given by

\begin{subequations}
   \begin{equation}
	\left( \partial_t - \mathcal{L}_{\beta} \right) \tilde{\gamma}_{ij} = -2 \alpha \tilde{A}_{ij}
    \end{equation}

    \begin{equation}
    \label{eq:phi_orig}
	\left( \partial_t - \mathcal{L}_{\beta} \right) \phi = -\frac{1}{6}\alpha K
    \end{equation}

   \begin{multline}
	\left( \partial_t - \mathcal{L}_{\beta} \right) K = -\gamma^{ij} D_j D_i \alpha + \frac{1}{3} \alpha K^2 + \\
		\alpha \tilde{A}_{ij} \tilde{A}^{ij} + 4 \pi \alpha (\rho + S)
    \end{multline}

   \begin{multline}
	\left( \partial_t - \mathcal{L}_{\beta} \right) \tilde{A}_{ij} = e^{-4\phi} \left[ -D_i D_j \alpha + \alpha \left(R_{ij} - 8\pi S_{ij} \right) \right]^{TF}\\
		+ \alpha \left(K \tilde{A}_{ij} - 2 \tilde{A}_{il} \tilde{A}^l_j \right)
    \end{multline}
\text{and}
    \begin{multline}
	\left( \partial_t - \mathcal{L}_{\beta} \right) \tilde{\Gamma}^i = -\partial_j \left(2\alpha \tilde{A}^{ij} + \mathcal{L}_{\beta} \tilde{\gamma}^{ij} \right), \\
    \end{multline}
\end{subequations}
 where $\tilde{\gamma}_{ij}$ is the conformally-related 3-metric,
 $\alpha$ is the lapse, $\phi$ is the conformal exponent, $K$ is the
 trace of the extrinsic curvature $K_{ij}$, $\tilde{A}_{ij}$ is the
 conformal traceless part of the extrinsic curvature, $R_{ij}$ is the
 Ricci tensor associated with the 3-metric of spacelike hypersurfaces
 $\gamma_{ij}$, and $\tilde{\Gamma}^{i}$ are the conformal connection
 functions. The matter source terms $\rho$, $S_i$, and $S_{ij}$ are
 the usual projections of the stress-energy tensor. Finally, $D_i$ is
 the covariant derivative operator associated with $\gamma_{ij}$, and
 $\mathcal{L}_{\beta}$ is the Lie derivative with respect to the
 shift, $\beta^i$. For calculation of the Lie derivative terms and for
 further details, see \cite{Baumgarte1999}.

In terms of these variables, the Hamiltonian and momentum constraints
are given by
   \begin{multline}
   \label{eq:haml}
	0 = \mathcal{H} = \tilde{\gamma}^{ij} \tilde{D}_i \tilde{D}_j e^{\phi} - \frac{ e^{\phi}}{8} \tilde{R}  \\
			+ \frac{e^{5\phi}}{8} \tilde{A}_{ij} \tilde{A}^{ij} - \frac{e^{5\phi}}{12} K^2 + 2\pi e^{5 \phi} \rho
    \end{multline}
 and   
   \begin{equation}
	0 = \mathcal{M}^i = \tilde{D}_j \left(e^{6\phi} \tilde{A}^{ji} \right) - \frac{2}{3} e^{6 \phi} \tilde{D}^i K - 8 \pi e^{6 \phi} S^i.
    \end{equation}

In the continuum limit, both $\mathcal{H}$ and $\mathcal{M}^i$ must be
zero for the solution to be physical. Thus, during the numerical
evolution of the BSSN equations, these quantities must show convergence to
zero, in order to remain consistent with the Einstein equations.  We
monitor the $L_2$ norm of the Hamiltonian constrain,
$||\mathcal{H}||$, as defined e.g. in \cite{Etienne2008}, to test for
convergence in our simulations. A 4th-order accurate finite difference
method for the BSSN equations is employed as described
in~\cite{Etienne2008}.

\subsection{Treatment of constraint damping}
\label{sec:damping}

In simulations involving strongly gravitating matter, the growth of
the Hamiltonian constraint can be minimized by the addition of damping
terms to the BSSN equations. For example, it has been found that
adding in such a damping term to the evolution equation for $\phi$ can
improve the numerical stability and accuracy of the code
\cite{Duez2003,Duez2004}. As in~\cite{Duez2003}, we add a multiple of
the Hamiltonian constraint to Eq.~(\ref{eq:phi_orig}) to damp
Hamiltonian constraint-violating degrees of freedom, according to
\begin{equation}
\label{eq:phi}
 	\left( \partial_t - \mathcal{L}_{\beta} \right) \phi = -\frac{1}{6}\alpha K + c_{H} \mathcal{H},
\end{equation}
where $c_{H}$ governs the amplitude of the damping. We note that
eq.~(\ref{eq:phi}) is very similar to one of the modified evolution
equations proposed, at the analytic level, in 
\cite{Yoneda2002} 
to
control constraint violations. This version of the evolution equation
for $\phi$ is the standard implementation within the code of
Refs.~\coderefs for damping the Hamiltonian constraint, and has been
adopted in all binary neutron star simulations with this code. The
principal part of the new term is $c_H \mathcal{H} \propto c_H
\tilde\gamma^{ij}e^\phi \partial_i \partial_j \phi $. When $c_H>0$, it
is a 2nd-order parabolic operator which parabolically diffuses the
Hamiltonian constraint. As such, it is subject to a Courant stability
condition of the form $c_H \Delta t/(\Delta x)^2 \le A$ (with $A$ of
order $1/6$ in 3+1 dimensional explicit time integrations in flat
spacetime), where $\Delta t$ is the time step and $\Delta x$ is the
grid spacing. Depending on the implementation of $c_H$, the damping on
different levels of the grid can vary due to different refinement
levels having different grid spacing, leading to potential
complications as we discuss in more detail below.  Understanding how
the prescription for this parabolic constraint damping term affects
the convergence of the code is one of the main goals of this paper.

  \begin{table*}
\centering
\begin{tabular}{llllllllllll}
\hline \hline
Name && Damping prescription && Parametrization &&   Resolutions &&  \\
\hline 
 \vspace{0.05cm} 
 $\Delta t$-scaled  &&  $c_{H} = 0.08 \Delta t$  &&	GPP   &&  LR, MR, HR  \\  \\ 

  Modified $\Delta t$-scaled &&  $c_{H} = 0.08 \left[ \Delta x_{\rm HR}  / \Delta x_{i} \right] \Delta t$ &&   GPP   &&  LR, MR, HR  \\  \\

 Constant $c_H$   &&  $c_{H}$=0.0045~km &&  GPP &&  LR, MR,  HR   \\ 
           	  		       	    &&     &&	PWP &&  LR, MR, HR  \\  
\hline
\end{tabular}
\caption{Summary of tests run. The resolutions correspond to finest-level grid spacings of $\Delta x_{\rm LR} = 195$~m, 
$\Delta x_{\rm MR} = 156 $~m, $\Delta x_{\rm HR} = 125$~m.  }
  \label{table:simulations}
\end{table*}

In \cite{Duez2003}, and more generally in the standard
implementation of eq.~(\ref{eq:phi}) in the code of Refs.~\coderefs,
 the damping coefficient is set as 
\begin{equation}
\label{eq:const_cH}
c_H = c_{H1} \Delta t,
\end{equation} 
where $c_{H1}$ is a constant. We refer to this prescription as 
``$\Delta t$-scaled damping."
The rationale behind this form is that
the stability condition from the parabolic operator now becomes
$c_{H1} (\Delta t/\Delta x)^2 = c_{H1} {\rm CFL}^2 \le A$, where ${\rm
  CFL}=(\Delta t/\Delta x)$ is the Courant factor. Therefore, by
choosing $c_{H1}$ one can satisfy the parabolic Courant stability
condition independently of the grid spacing, since in these
integrations the CFL factor is often also fixed.  
However, when both $c_{H1}$ and the CFL factor are fixed, the degree of 
damping (i.e., $c_H$) will decrease when moving
to evolutions with higher spatial resolution, since the time-step on a given
refinement level gets smaller as the overall resolution of the simulation
is increased.
 As we will show using a baseline set of simulations with a constant
$c_{H1}= 0.08$, this leads to undesirable behavior in the growth of
the Hamiltonian constraint, where the reduced damping leads to larger
constraint violations for evolutions at higher resolution.  The
$\Delta t$-scaled damping prescription is what has been used in all
prior simulations with the code of Refs.~\coderefs.

In an attempt to reduce this
behavior, we additionally implement a modified version of this
parabolic damping treatment, in which $c_{H1}$ is scaled by the global
resolution. In particular, we use the standard prescription of
eq.~(\ref{eq:const_cH}) for our highest resolution evolution, while each lower resolution evolution, $i$, uses a scaled damping according to 
\begin{equation}
\label{eq:scaled_cH}
c_H = c_{H1} \left( \Delta x_{\rm HR} / \Delta x_i \right) \Delta t.
\end{equation}
We note that the index $i$ above does not label refinement levels, but rather 
simulations with the same grid hierarchy and different spatial
resolution.  We scale the damping coefficient with respect to the
highest resolution performed, so that the Courant stability condition
is always satisfied.  For the three resolutions used in this work
(which vary in grid spacing by factors of 1.25; see
Sec.~\ref{sec:setup}), the scaled coefficients are thus 0.0512, 0.064,
and 0.08.  In this prescription, which we refer to as ``modified
$\Delta t$-scaled damping", the degree of damping still varies across
the refinement levels (due to the dependence of $c_H$ on $\Delta t$),
but the amount of damping on a \textit{given} level is the same for
evolutions with different resolutions.

Both of these prescriptions suffer from discontinuities in the damping
between refinement levels with different timesteps $\Delta t$.  Such
discontinuities may become problematic when integrating global
quantities or as waves propagate across the grid. For example, in a
recent study, Ref.~\cite{Bozzola2021} found that requiring the
Kreiss-Oliger dissipation to be continuous everywhere on the grid
(i.e., eliminating the discontinuities of the standard prescription)
was essential to achieving long-term, stable evolutions of charged
binary black holes.

Motivated by this finding, we thus also implement a third
prescription, which ensures a continuous parabolic damping operator
for the Hamiltonian constraint. We define this case simply as
\begin{equation}
c_H = c_{H1},
\end{equation}
where we use the finest-time level of our highest-resolution grid
($\Delta t = 0.05625$~km) to set $c_H = 0.0045$~km in the simulations
presented below. This choice ensures that the Courant stability
condition is satisfied on all refinement levels for all resolutions
explored in this work.  Additionally, because $c_H$ no longer scales
with the local timestep, the damping is the same everywhere on the
grid, thereby ensuring that the same set of partial differential
equations will be solved across all refinement levels and
independently of the timestep. In other words, this is the only
prescription for which the continuum formulation is fixed
independently of the grid.  We refer to this final prescription as
``constant $c_H$ damping."

\subsection{Equation of state parametrization}
\label{sec:eos}

While discontinuities in constraint damping can cause numerical issues
when integrating quantities across the grid or as waves propagates across refinement
levels, discontinuities in the sound speed resulting from the equation
of state (EoS) can introduce unphysical solutions in the hydrodynamics
of the evolution. This, in turn, may also affect the convergence properties of
the simulation.  For example, in the context of spectral/finite-volume
methods, it has been shown that discontinuities in the sound speed can
lead to increased phase errors in the inspiral of a binary neutron
star merger simulation, compared to results found with a smooth EoS
\cite{Foucart2019}. This is of particular interest, given the
prevalence in the merger literature of the piecewise polytropic method
for modeling the EoS \cite{Ozel2009,Read2009}, in which the EoS is
non-smooth. This leads us to the second main goal of the present
paper: to investigate how the smoothness of the EoS affects the
convergence of a binary neutron star merger simulation in the context
of finite-difference/volume methods. The finite-volume method adopted
in this work uses the Harten-Lax-Van Leer scheme in conjunction with
the piecewise parabolic reconstruction method as described
in~\cite{PhysRevD.85.064029,PhysRevD.85.024013}.

To that end, we perform simulations with two different EoS
parameterizations: standard piecewise
polytropes~\cite{Ozel2009,Read2009} and the generalized piecewise
polytrope (GPP) framework introduced recently by \cite{OBoyle2020}.
The GPP parametrization was designed to share the accuracy 
and flexibility of the PWP
approach, while also ensuring a smooth (i.e., continuous and
differentiable) pressure function. In this section, we briefly
summarize the two approaches.

For a PWP EoS, the pressure between two dividing densities,
$\rho_{i-1}$ and $\rho_{i}$, is defined as
\begin{equation}
P(\rho) = K_i \rho^{\Gamma_i}, 		\quad  \rho_{i-1} < \rho \le \rho_{i}
\end{equation}
where the polytropic constant, $K_i$, is determined by imposing continuity in the pressure
 between adjacent polytropic segments, according to
\begin{equation}
K_i = \frac{P_{i-1}}{\rho_{i-1}^{\Gamma_i} } =  \frac{P_{i}}{\rho_{i}^{\Gamma_i }}.
\end{equation}
The polytropic index, $\Gamma_i$, is then given by
\begin{equation}
\label{eq:gamma_pwp}
\Gamma_i \equiv \frac{\partial \ln P}{\partial \ln \rho} = \frac{\log{ \left( P_i / P_{i-1}\right)}}{ \log{\left(\rho_i / \rho_{i-1} \right)}}.
\end{equation}
In this formulation, it is clear that the pressure is continuous but not differentiable. 

In the generalized PWP 
formulation of \cite{OBoyle2020}, the pressure is instead defined as
\begin{equation}
\label{eq:P_GPP}
P(\rho) =  \overline{K}_i \rho^{\overline{\Gamma}_i} + \Lambda_i, 		\quad  \rho_{i-1} < \rho \le \rho_i
\end{equation}
where $\overline{\Gamma}_i \equiv \partial \ln P / \partial \ln \rho$ is the new adiabatic index, which now differs
from the right-hand side of eq.~(\ref{eq:gamma_pwp}). Imposing \textit{differentiability} requires the polytropic constant to be 
\begin{equation}
\label{eq:K_GPP}
\overline{K}_i  = \overline{K}_{i-1} \left( \frac{\overline{\Gamma}_{i-1}}{{\Gamma}_i} \right) \rho_{i-1}^{\overline{\Gamma}_{i-1} - \overline{\Gamma}_i } 
\end{equation}
Finally, the new parameter, $\Lambda_i$, is introduced to impose continuity in the pressure, such that 
\begin{equation}
\label{eq:lambda_GPP}
\Lambda_i = \Lambda_{i-1} + \left( 1 - \frac{\overline{\Gamma}_{i-1}}{\overline{\Gamma}_i } \right ) \overline{K}_{i-1} \rho_{i-1}^{\overline{\Gamma}_{i-1} }.
\end{equation}
With this construction, the pressure is continuous and differentiable,
ensuring that the sound speed will also be continuous
\cite{OBoyle2020}.

In this paper, we adopt the nuclear EoS ENG, which predicts a radius of 12.06~km for a 1.4~$M_{\odot}$ cold,
non-rotating neutron star, and a corresponding maximum mass of 2.24~$\Ms$ \cite{Engvik1994,Engvik1996}. 
These properties are both consistent with latest astrophysical constraints 
\cite{Ozel2016,Antoniadis2013,Fonseca2016,Cromartie2020,Baiotti2019,Raithel2019a,Chatziioannou2020,Annala:2021gom}.
 For the PWP case,
we use the coefficients provided by \cite{Read2009} for the three-polytrope approximation of
ENG at high-densities. We additionally use the PWP coefficients from that paper for
 the SLy EoS \cite{Douchin2001} to describe the low-density crust.
  For the GPP formulation, the parameters cannot be calculated directly, but rather must
 be fit numerically. To do so, we use the fit coefficients for SLy provided by \cite{OBoyle2020}
 to describe the low-density crust. We then perform a Markov Chain Monte Carlo simulation 
 to find the best-fit parameters for a three-polytrope approximation to ENG at higher densities.
We report the details of this fit and compare the results to the PWP approximation in Appendix~\ref{sec:EOSfit}.

 \begin{figure*}[!ht]
\centering
 \includegraphics[width=\textwidth]{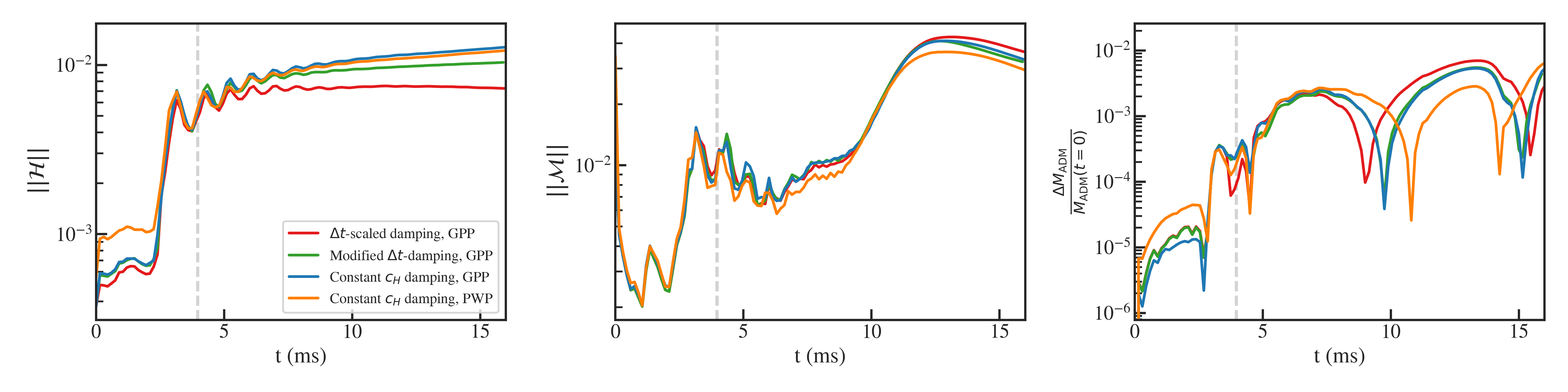}
 \caption{\label{fig:conservedQ} Comparison of global properties for
  the four cases studied in this paper. From left to right: the $L_2$
  norm of the Hamiltonian constraint, the $L_2$ norm of the momentum
  constraint, and the fractional change to the ADM mass. For the
  latter quantity, we compute the fractional change with respect to
  $M_{\rm ADM}$ at $t=0$, accounting for the losses due to
 gravitational waves leaving the computational domain.  All
  quantities are plotted from the medium resolution evolutions. The
  vertical dashed line indicates the time of merger, which corresponds
  to the time of maximum gravitational wave strain.  By reducing the
  damping parameter $c_H$ to its smallest value to make it continuous
  across the grid (i.e., constant $c_H$ damping, shown in
  blue and orange), the Hamiltonian constraint
  violations grow to slightly larger values over time.  The momentum
  constraint violations and ADM mass conservation are comparable for
  all three damping prescriptions and for both the PWP and GPP EoSs. }
\end{figure*}

\subsection{Initial conditions and numerical setup}
\label{sec:setup}
For all simulations, we construct binary initial data with LORENE \cite{Lorene}.
The initial configurations describe
two unmagnetized, irrotational, equal-mass neutron stars in a quasi-circular orbit, with an 
Arnowitt-Deser-Misner (ADM) mass of $\sim2.76~\Ms$ and an initial separation of 32.5~km.
The neutron stars start at zero-temperature and are described by either the PWP representation
of the ENG EoS \cite{Engvik1994,Engvik1996,Read2009}, or the generalized PWP approximation
 of ENG described further in Appendix ~\ref{sec:EOSfit}.

 We extend the cold parametric EoSs to finite temperatures
 using the $M^*$-framework of \cite{Raithel2019}, which is based on a two-parameter
 approximation of the particle effective mass and includes the leading-order 
 effects of degeneracy in the thermal prescription. The implementation and validation
 of this framework into our code was recently described in \cite{Raithel2021}.
In this work, we use an intermediate set of $M^*$-parameters to describe
the matter in the degenerate regime, corresponding to $n_0=0.12$~fm$^{-3}$ and
$\alpha = 0.8$. For additional details, see \cite{Raithel2019, Raithel2021}.

For each binary evolution, we use nine spatial refinement levels, which are
separated by a 2:1 refinement ratio. We additionally use six temporal
refinement levels, such that the coarsest four levels are evolved with identical
timesteps, while the remaining levels are separated by a 2:1 ratio.
The CFL factor on the finest refinement level is 0.45. 
The outer boundary of the
computational domain is located at 3200~km, and 
we impose equatorial symmetry to reduce computational costs.
 We study three
different resolutions, with grid spacings on the finest level of
$\Delta x_{\rm LR} = 195$~m, $\Delta x_{\rm MR} = 156.25$~m, and
$\Delta x_{\rm HR} = 125$~m, where the subscripts indicate low,
medium, and high resolution, respectively.  These grid spacings
correspond to $\sim$100, 125, and 156 points across the diameter of
each initial neutron star in the x-direction (i.e., along the line
  connecting the two stars). 

\subsection{Summary of simulations performed}
All together, we perform a total of 4 new convergence studies.  This
includes three convergence studies to study the three parabolic
constraint damping approaches described in Sec.~\ref{sec:damping}. For each
of these studies, we use the new GPP EoS parametrization, in order to
validate its implementation into our code. We additionally
perform a fourth study with the PWP EoS parametrization together with
the continuous parabolic constraint damping which, as we will show,
achieves the best convergence properties of the various damping
prescriptions studied.  This fourth study enables us to compare the
impact of the EoS parametrization on the convergence properties,
starting from a robust baseline. We summarize these tests in
Table~\ref{table:simulations}.

\section{Results}
\label{sec:results}

We now turn to the results of the simulations for each of the three
parabolic damping treatments, as well as for the two EoS parametrizations
studied in this work. In all cases except one, we evolve the mergers
from approximately the final two orbits, through the merger itself,
and for $\sim10-20$~ms post-merger, with the longer range of
evolutions possible for the lower resolutions studied. The one
exception is the PWP evolution at the highest resolution studied,
which we evolve only in the inspiral phase, as this was 
sufficient for the comparisons we will make below.

\subsection{Comparison of global quantities}
\label{sec:conserved}
In order to confirm the stability of the evolutions, 
we start by comparing their global spacetime properties, including the $L_2$
norms of the Hamiltonian constraint, $||\mathcal{H}||$, and momentum constraint,
$||\mathcal{M}||$ (for
definitions, see eqs. 40-43 of \cite{Etienne2008}), as well as the ADM
mass. We show the evolution of these three quantities for all four of 
our tests in Fig.~\ref{fig:conservedQ}. All results correspond to the
evolutions at our intermediate resolution. The Hamiltonian constraint
jumps at merger, as is commonly found in compact binary mergers, and
continues to slowly grow at late times. The magnitude of this late-time
growth is directly governed by the degree of damping in eq.~(\ref{eq:phi}).
For example, for the case of the $\Delta t$-scaled damping, 
the damping is enhanced on refinement levels with larger timesteps; thus,
there is overall more damping in this prescription, and the growth of the
Hamiltonian constraint is correspondingly reduced. On the other hand,
the constant $c_H$ prescription has the smallest degree
of damping ($c_H = 0.0045$~km on \textit{every} refinement level),
and thus the Hamiltonian constraint exhibits slightly larger growth over
time.

We additionally find some difference in $||\mathcal{H}||$ during the
inspiral between the PWP evolution (which uses the constant $c_H$
prescription) and all of the GPP evolutions, with the PWP result being
$\sim50\%$ larger.  This difference enters already at $t=0$,
suggesting that the difference in the smoothness of the sound speed
between these parametrizations affects the construction of
the initial data.

The momentum constraint violations, $||\mathcal{M}||$, shown in the
middle panel of Fig.~\ref{fig:conservedQ}, are comparable for all
tests considered here.  Finally, the right panel of
Fig.~\ref{fig:conservedQ} shows the ADM mass ($M_{\rm ADM}$), i.e.,
the contribution to the ADM mass integral in our computational domain,
compared to the $t=0$ value for each test, with losses due to
gravitational waves added back in . Overall, we find small errors
$\lesssim0.7\%$ in the ADM mass conservation over the course of the
simulation. We find no significant differences in the conservation of
$M_{\rm ADM}$ between the different parabolic constraint damping
treatments. Finally, in comparing between the two EoS
parametrizations, we again find worse conservation of $M_{\rm ADM}$
for the PWP parametrization during the inspiral. However, at late
times, the conservation is marginally better for the PWP
 parametrization and, overall, we conclude that the difference between
the conservation of $M_{\rm ADM}$ for the PWP and GPP parametrizations
is small.

All together, these results confirm that the three parabolic damping
treatments and two EoS parametrizations studied in this paper can be
used to stably evolve binary neutron stars, with robust preservation
of the global spacetime properties reported here.

\subsection{Convergence of the Hamiltonian constraint}
\label{sec:conv}

 \begin{figure*}[!ht]
\centering
\includegraphics[width=\textwidth]{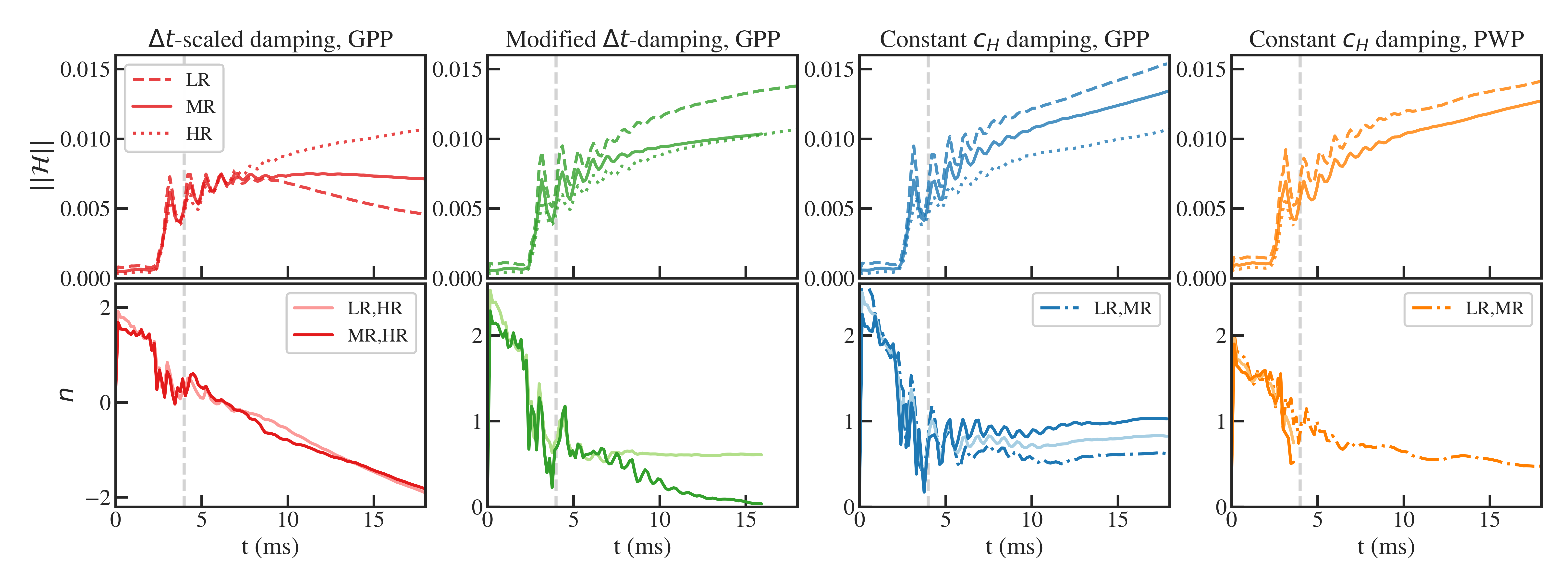}
\caption{\label{fig:Haml} Top panel: $L_2$ norm of the Hamiltonian constraint
   for each of the three parabolic constraint damping
  treatments and two parametrizations studied in this work. From left
  to right, we show the results for the $\Delta t$-scaled damping, the
  modified $\Delta t$-scaled damping, and the constant $c_H$ damping
  prescription, respectively. Bottom panel:
  Convergence order, $n$, for the Hamiltonian constraints, assuming
  $||\mathcal{H}||$ approaches zero for infinite resolution. The left three
  columns all use the GPP parametrization, while the right-most column
  corresponds to the PWP parametrization.  Only with the constant
  $c_H$ treatment do we find positive convergence at late
  times post-merger, across all resolutions studied.} 
\end{figure*}

Although we find similar conservation of the Hamiltonian constraint for
 the various parabolic constraint damping
treatments and EoS parametrizations, we find significant differences
in the numerical \textit{convergence} of $||\mathcal{H}||$ between these treatments.

We show the evolution of $||\mathcal{H}||$ at all three resolutions
for each of our studies in the top row of Fig.~\ref{fig:Haml}. The bottom
row shows the corresponding convergence order, assuming $||\mathcal{H}||$ 
approaches zero with increasing resolution. This convergence order is defined as
\begin{equation}
n = \log \left( \frac{ ||\mathcal{H}(\Delta x_1)|| }{ ||\mathcal{H}(\Delta x_2)|| } \right) / 
	\log \left( \frac{ \Delta x_1}{ \Delta x_2 } \right),
\end{equation} 
for two grid spacings, $\Delta x_{1,2}$.

In general, we find similar qualitative behavior in the growth of
$||\mathcal{H}||$ at all resolutions to what was described for the
intermediate resolution in Sec.~\ref{sec:conserved}.  The one notable
exception is the low-resolution evolution with $\Delta t$-scaled damping
 ($c_H = 0.08 \Delta t$; red dashed line in
Fig.~\ref{fig:Haml}), which shows a reduction in $||\mathcal{H}||$ at
late times.  The reason for this turnover is related to the fact that the 
low-resolution evolution has the largest timesteps,
$\Delta t$ (located on the lowest-resolution refinement levels).
Because $c_H$ scales with $\Delta t$ for this prescription, this
leads to the highest degree of damping and, hence, the
late-time decrease in $||\mathcal{H}||$.

The $\Delta t$-scaled damping
prescription also exhibits a troublesome trend, in which
$||\mathcal{H}||$ in the highest resolution evolution actually
\textit{exceeds} the value at lower resolutions, starting shortly
after merger. This is due to the fact that at higher resolutions,
 $\Delta t$ becomes smaller (for fixed CFL number) and 
 hence the damping also becomes
smaller in this prescription. This results in a negative convergence
order, as shown in the bottom, left panel of Fig.~\ref{fig:Haml}.
These findings are consistent with our previous results from
\cite{Raithel2021}, in which we used the same $c_H = 0.08 \Delta t$ damping
 prescription and
similar initial conditions, but with the standard PWP parametrization
of ENG.

Our second parabolic constraint-damping prescription is scaled by the
resolution such that $c_H = 0.08 \left( \Delta x_{\rm HR} / \Delta
x_{i} \right) \Delta t$. Although $c_H$ still scales with the local
timestep, and thus can vary depending on the refinement level, this
prescription ensures that the amount of damping on a \textit{given}
refinement level is the same for each of the three resolutions studied
in this work.  The results from evolutions with this treatment, which
start from identical initial data as the previous case, are shown in
the second column of Fig.~\ref{fig:Haml}.  With this modified $\Delta
t$-scaled damping, we find improved convergence behavior, with
positive (but decaying) convergence order achieved for nearly 10~ms
post-merger. However, at later times $(t - t_{\rm merger} > 12$~ms),
the convergence order again becomes negative between the two highest
resolutions studied, indicating loss of convergence.

Finally, the right two columns of Fig.~\ref{fig:Haml} show the results
from the evolutions with the constant $c_H = 0.0045$~km damping, for
both the GPP and PWP parametrizations.  This prescription ensures
equal damping on different refinement levels, so that the parabolic
damping operator added to the BSSN equations is continuous across the
entire grid.  With this continuous damping prescription, we find a
positive and stable convergence order until the end of our simulations
($\sim 15$~ms post-merger), for either EoS parametrization. In
particular, the medium and high resolution GPP evolutions indicate
convergence-to-zero at approximately first order.  This is consistent
with the order of our hydrodynamical numerical scheme which becomes
first-order accurate when shocks arise, e.g., post-merger. We note
that the convergence order is lower when calculated with the
low-resolution data ($n\approx 0.5-0.6$), which suggests that the
lowest resolution in our study is not be in the convergent regime for
all times, despite covering the diameter of each initial neutron star
with $\sim 100$ grid points. Nevertheless, the finding of $n_{\rm
  MR,HR} \approx1$ convergence only with the constant $c_H$
prescription provides strong evidence that using a constant parabolic
damping parameter is necessary to achieve the expected convergence
order.

In other words, we find that discontinuities in the
parabolic damping term in eq.~(\ref{eq:phi}) between adjacent
refinement levels lead to a decay in the convergence of
$||\mathcal{H}||$. The reasons for this decay are twofold.  First, if
there are different degrees of damping on different refinement levels,
the set of partial differential equations that is being
solved will also change depending on the level. In addition, the
discontinuities at the refinement level boundaries may propagate in
ways that further spoil the numerical convergence. These effects
accumulate when integrating $\mathcal{H}$ across the grid to compute
the $L_2$-norm and the convergence order is correspondingly reduced.

We also compare between the PWP and GPP parametrizations of the EoS in
the right two columns of Fig.~\ref{fig:Haml}.  For both
parametrizations, we use the constant $c_H$ damping, to ensure
optimal convergence in the comparison.  We find similar convergence
results between the two cases, with small differences in the initial
convergence order.  In particular, we achieve a slightly higher
initial convergence order, $n\simeq 2.25$ (2.50), with the GPP
parametrization, compared to $n\simeq 1.90$ (1.96) for the standard
PWP parametrization, for the convergence order calculated between the
high and medium (low) resolution evolutions.  As discussed in
Sec.~\ref{sec:conserved}, where we found larger $||\mathcal{H}||$ at
$t=0$ for the PWP parametrization, the slight reduction in the initial
convergence order seems to originate at the initial data level.  For
both parametrizations, the post-merger convergence order asymptotes to
$n\simeq0.5-0.6$, calculated between the low and intermediate
resolutions. Given the similarity between the results at low and medium
resolutions for the GPP and PWP parametrizations, we did not evolve
the highest resolution past the merger for the PWP case. 
Thus, in summary, the PWP and GPP parametrizations lead to overall
similar convergence of $||\mathcal{H}||$. 

Finally, we note that it is possible that the Kreiss-Oliger
dissipation, which is added (at fifth order) to all evolved BSSN and
gauge variables at the refinement level boundaries and which has a
diffusive coefficient that is different on different refinement
levels, may also suffer the same issues as the parabolic term in
eq.~(\ref{eq:phi}). Implementing a continuous prescription for the
Kreiss-Oliger dissipation may further improve the numerical
convergence of our results. 

In summary, we find that a continuous parabolic-constraint-damping
prescription is necessary to recover convergence of $||\mathcal{H}||$
in the post-merger phase.  We note that this conclusion was exhibited
in the context of parabolic constraint damping in the BSSN
formulation, but it should be the case for more general types of
parabolic constraint damping at the analytic level of formulations of
the Einstein
equations~\cite{Paschalidis:2007ey,Paschalidis:2007cp}. However, this
is not a problem for low-order constraint damping terms such as those
of the conformal Z4 \cite{Bernuzzi2010,Hilditch2013,Alic2013} or the
generalized harmonic \cite{Gundlach2005,Pretorius2005,Lindblom2006}
families of formulations. It is not obvious how, or to what extent,
discontinuities across refinement levels in Kreiss-Oliger dissipation
operators might affect such formulations.  We leave the investigation
of this question to future work.

\subsection{Comparison of gravitational wave signals}
\label{sec:gw}
\begin{figure}[!ht]
\centering
\includegraphics[width=0.45\textwidth]{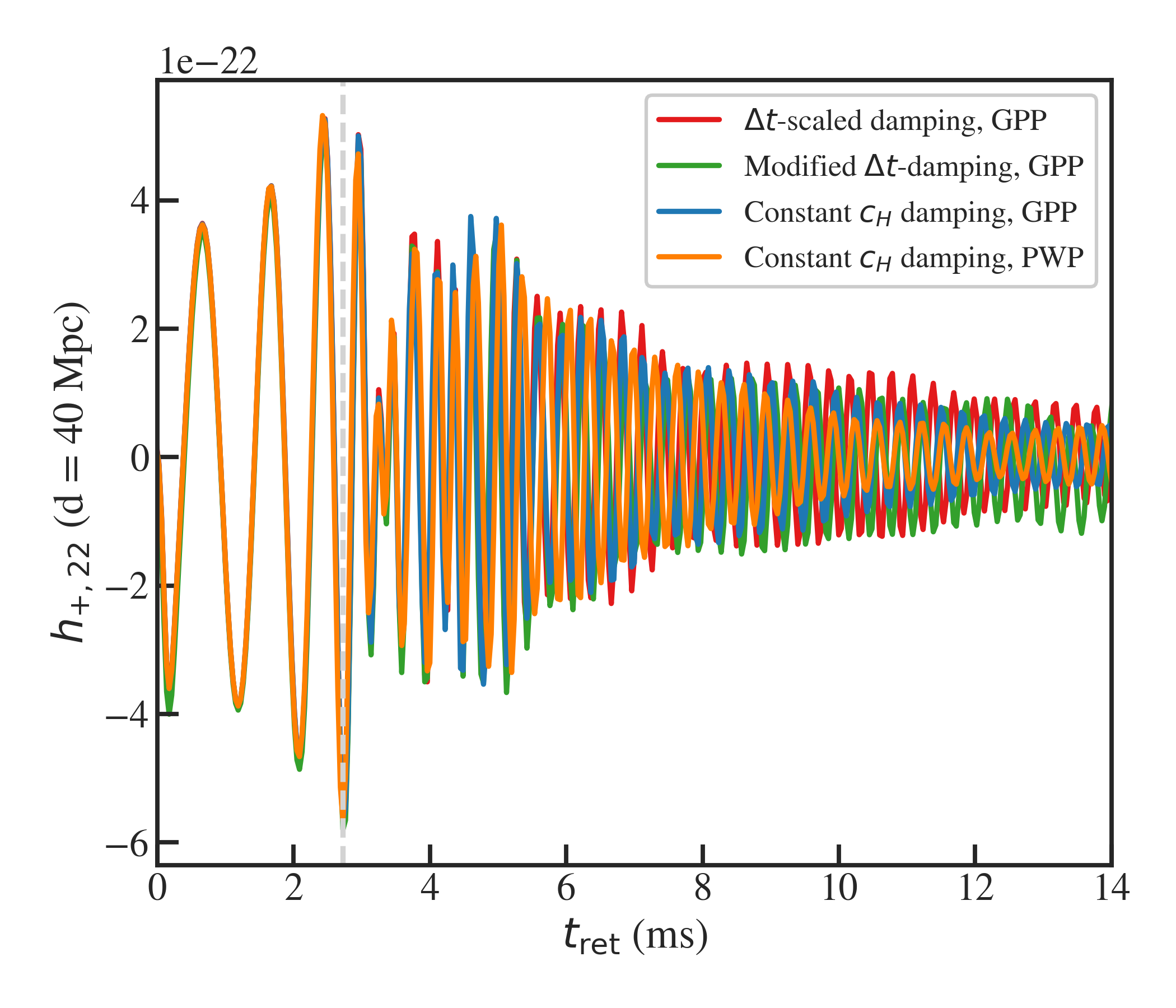}
\caption{\label{fig:strain} The $\ell=m=2$ mode of the plus-polarized
  gravitational wave strain for a merger at a distance of 40~Mpc, for
  each of the parabolic constraint damping treatments and EoS parametrizations considered
  in this work. The strain is plotted as a function of the retarded
  time.  All results are shown for the medium resolution simulations.
  The inspirals in all cases are nearly identical, but small
  differences emerge in the post-merger phase.}
\end{figure}

Finally, we turn to the gravitational wave signals extracted from each of our
simulations. 
We show the $\ell=m=2$ mode of the plus-polarized
gravitational wave strain, $h_{2,2}^+$, for each of our simulations in
Fig.~\ref{fig:strain}.  Theses signals correspond to face-on mergers,
located at a distance of 40~Mpc.  We find that the gravitational wave
signals leading up to merger are nearly identical for all parabolic constraint damping
treatments and EoS parametrizations considered in this work. However,
small differences emerge following the merger. We will spend the
remainder of this section discussing these differences in detail.  For
the specifics of the analysis methods used to extract the
gravitational wave strains and to calculate the corresponding spectra
for our simulations, see Appendix~C of \cite{Most2021}.

 \begin{figure*}[!ht]
\centering
\includegraphics[width=\textwidth]{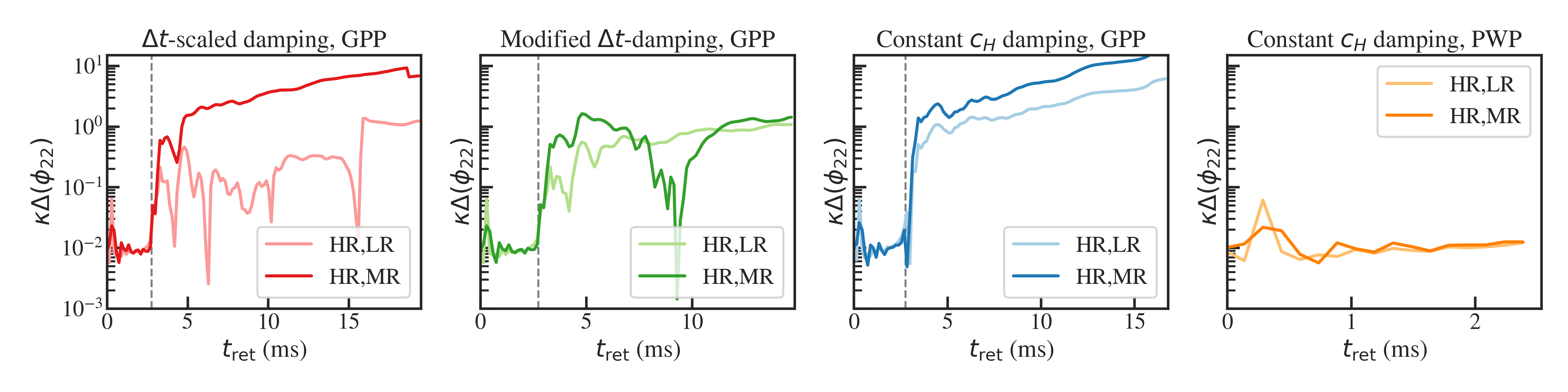}
\caption{\label{fig:conv_phi} Self-convergence of the phase of the
  $\ell=m=2$ mode of the gravitational wave signal, for (from left to
  right) the GPP parametrization with $\Delta t$-scaled damping, the GPP
  parametrization with the modified $\Delta t$-scaled damping, the GPP and PWP parametrizations
  with constant $c_H$ damping, respectively. All results are scaled for second-order
  convergence, according to eq.~(\ref{eq:selfConv}). }
\end{figure*}

 Before discussing these signals in detail, we first briefly comment on
the convergence properties of the gravitational wave strain.  
We find approximately second-order
self-convergence in the phase of the gravitational wave strain,
$\phi_{2,2}$, during the inspiral, for all four of our tests. Second-order
self-convergences requires that
\begin{equation}
\frac{\phi_{\rm LR} - \phi_{\rm HR}}{ \phi_{\rm MR}- \phi_{\rm HR}} = 
\frac{ \left( \Delta x_{\rm LR}/\Delta x_{\rm HR} \right)^2-1}{\left( \Delta x_{\rm MR}/\Delta x_{\rm HR} \right)^2-1},
\end{equation}
where we have suppressed the $(2,2)$ subscript on $\phi$ for clarity.
We can rearrange this expression to alternatively obtain
\begin{multline}
\label{eq:selfConv}
\left( \phi_{\rm LR} - \phi_{\rm HR} \right) \left[ \left( \Delta x_{\rm MR}/ \Delta x_{\rm HR} \right)^2-1 \right] = \\
\left( \phi_{\rm MR}-\phi_{\rm HR} \right) \left[ \left(  \Delta x_{\rm LR}/ \Delta x_{\rm HR} \right)^2-1 \right].
\end{multline}
We plot these scaled, differential quantities in
Fig.~\ref{fig:conv_phi}, where $\kappa \Delta \phi_{2,2}$ corresponds
to the left- or right-hand side of eq.~(\ref{eq:selfConv}), depending
on the color of the line.  The alignment of the scaled $\Delta
\phi_{2,2}$ curves during the inspiral indicates that second-order
self-convergence is indeed achieved.  However, after the merger, the
self-convergence is lost in all four cases, regardless of the damping
treatment or EoS parametrization.  Although the curves in
Fig.~\ref{fig:conv_phi} look very different during the post-merger
evolution, in all four cases the order of self-convergence becomes
negative at late times. As a result, the late-time differences in
Fig.~\ref{fig:conv_phi} between damping treatments do not hold much
significance. We find similar results for the amplitude of the
  $h_{2,2}$ and the instantaneous gravitational wave frequency, as
  well as for other quantities, such as the maximum
    rest-mass density and the maximum conformal exponent $\phi$. That
    is, for all of these quantities, we find approximate, second-order
    self-convergence during the inspiral, but that the
    self-convergence disappears at late times post-merger.

The lack of self-convergence at late times may be a consequence of the
lowest resolution evolution not being in the convergent regime (see
Sec.~\ref{sec:conv}), or it may have another origin. In either case,
without this convergence, we cannot draw strong conclusions about the
differences in the waveforms seen in Fig.~\ref{fig:strain}.  For
example, Fig.~\ref{fig:strain} indicates a possible difference in the
damping rate of the post-merger gravitational wave amplitudes,
depending on the parabolic damping treatment that is adopted. However,
whether this behavior is physical or numerical in nature requires further
investigation.

\begin{figure}[!ht]
\centering
 \includegraphics[width=0.45\textwidth]{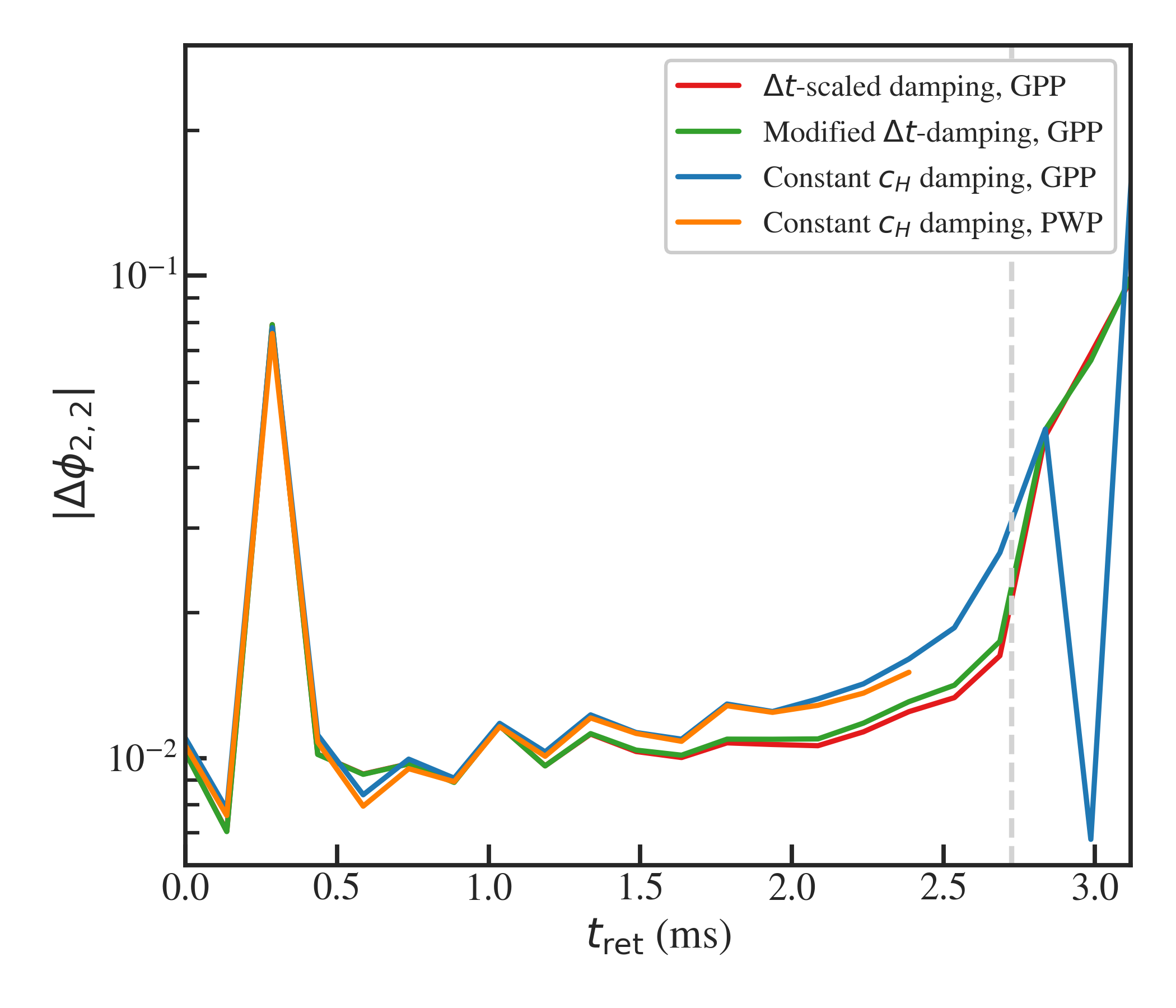}
 \caption{\label{fig:phaseDiff} Difference in the gravitational wave
  phase, $\phi_{2,2}$, between the phase extracted from the highest
  resolution evolution and the Richardson extrapolation of the phase, assuming second-order
  convergence during the inspiral. The standard and modified $\Delta t$-scaled
   parabolic damping
  treatments lead to similar phase errors during the inspiral, while
  we find slightly larger phase errors for the constant $c_H$ prescription.  The
  phase errors are nearly identical for the PWP and GPP
  parametrizations.}
\end{figure}

Nevertheless, we can take advantage of the approximate second-order
self-convergence during the inspiral, to calculate the Richardson
extrapolation of the inspiral phase using the low and high
resolutions. We then estimate the numerical errors during the inspiral
as the difference between this Richardson extrapolation and the phase
extracted from our highest resolution evolution.  We show the
resulting error etimates, $\Delta \phi_{2,2}$, in
Fig.~\ref{fig:phaseDiff}.

We find that the inspiral phase errors are slightly larger for the
evolutions with constant $c_H$ damping, but that, over the
course of the (relatively short) inspiral, the differences between the
various parabolic damping prescriptions are generally small.

Interestingly, we do not find significant differences in the phase
errors between the GPP and PWP parametrizations. This is in contrast
to a previous study by \cite{Foucart2019}, which showed that a
continuous sound speed significantly reduced the inspiral phase
errors, compared to a piecewise polytropic approximation of the same
EoS. The authors in that study evolved the spacetime using spectral
methods, and used a spectral expansion of the EoS to ensure smoothness
(which has a different construction than the GPP approximation used
here), and they also used a different nuclear EoS (SLy). Additionally,
in that study, the authors start their simulations at larger initial
binary separation than we do, thereby allowing more gravitational wave
cycles for differences in phase to accumulate.  Any of these reasons
could explain the smaller effect that we find here. We leave further
study of this point to future work.

As a further check on our results, we also compute the approximate
errors during the inspiral for the maximum rest-mass density,
$\rho_{b,\rm max}$, and the maximum conformal exponent, $\phi_{\rm
  max}$.  The calculation is identical to that described above, and we
find errors in these quantities that are similar to what is shown in
Fig.~\ref{fig:phaseDiff}. In particular, we find that the approximate
(Richardson-extrapolated) errors in $\rho_{b,\rm max}$ and $\phi_{\rm
  max}$ are comparable for each of the the different parabolic-damping
treatments as well as for the two EoS parametrizations. In other
words, both the self-convergence properties and the approximate
inspiral errors for $\rho_{b,\rm max}$ and $\phi_{\rm max}$ are
similar to what is reported here for the gravitational waves.

 \begin{figure}
\centering
 \includegraphics[width=0.45\textwidth]{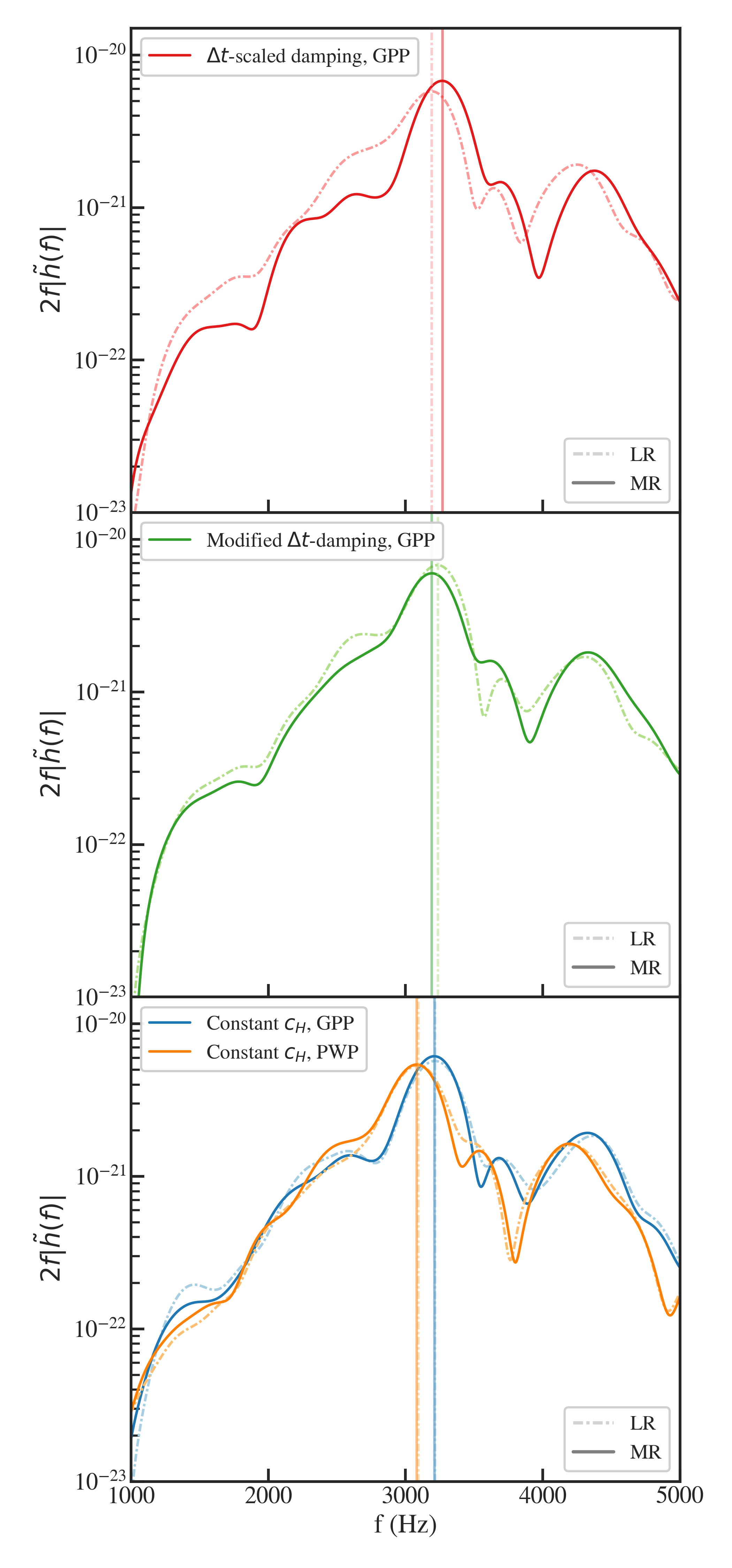}
 \caption{\label{fig:hc_multiRes} Characteristic strain for each of the
  parabolic damping treatments and EoS parametrizations considered in this
  work, for a face-on merger at 40~Mpc. These spectra include all
  $\ell=2,3$ modes of the gravitational wave strain. The top panel
  shows the results for the standard $\Delta$t-scaled damping;
  the middle panel shows the results for the modified $\Delta t$-scaled
  damping, in which the
  damping coefficient is scaled by the global resolution; and the
  bottom panel shows the results for the evolutions with constant $c_H$ damping, 
  for both the GPP (blue) and PWP (orange) parametrizations
  of the EoS. In all cases, the low resolution results are shown in
  the light dash-dot lines , while the medium resolution results are
  shown as the darker, solid lines. }
\end{figure}

In order to compare the gravitational waves in the post-merger phase, we compute the characteristic strain
for each evolution, according to
\begin{equation}
h_c = 2 f \tilde{h}(f)
\end{equation}
where $f$ is the frequency and $\tilde{h}(f)$ represents the Fourier
transform of the strain. To compute $\tilde{h}(f)$, we first window
each signal between $t=t_{\rm merger}$ and 12~ms post-merger. This
window length corresponds to our shortest-duration evolution, and
ensures that all spectra have the same effective
resolution. Additionally, we include all $\ell=2,3$ modes in these
spectra and and assume a face-on merger located at 40~Mpc.  For
additional details on our calculation of $h_c$, see Appendix C of
\cite{Most2021}.  We show the resulting spectra in
Fig.~\ref{fig:hc_multiRes}.

In general, we find that the post-merger spectra roughly agree with
one another, for all four tests considered.  However, we find a few
notable differences. First, we find a small difference in the
approximate errors in the peak frequency, $f_{\rm peak}$, of the
spectra, for the different parabolic damping treatments.  We estimate
the errors ($\sigma_f$) in $f_{\rm peak}$ simply as the difference
between the low and medium resolutions; this provides a rough estimate
of the numerical error. We find that the error is largest for the
non-continuous parabolic damping treatments: $\sigma_f \approx 80$ and
45~Hz for the standard and modified $\Delta t$-scaled damping
treatments, respectively. These differences correspond to fractional
errors of $\sim$2\% and 1\%, calculated with respect to the medium resolution. 
In contrast, the errors are significantly reduced with the constant 
$c_H$ damping:  we find $\sigma_ f
\lesssim 5$ Hz (0.2\%) and 10~Hz (0.4\%) with this damping prescription
evolved with either the GPP or PWP
parametrization, respectively.  This is consistent with the overall
worse convergence behavior found for $||\mathcal{H}||$ with the
discontinuous parabolic damping prescriptions (see
Sec.~\ref{sec:conv}).  In other words, we find that using a continuous
damping helps to reduce the approximate errors in the post-merger
spectra, at least as defined here.

We find that $f_{\rm peak}$ agrees to within this approximate error
for all three damping treatments with the GPP EoS parametrization. In
particular, $f_{\rm peak}\simeq3200$~Hz in all three cases, suggesting
that the details of the parabolic damping method do not significantly
influence the peak frequency of the post-merger gravitational waves.

In contrast, we find a systematic difference of $\sim 130$~Hz 
(fractional difference of 4\%)
in $f_{\rm peak}$ between the GPP and PWP parametrizations with the
continuous $c_H$ prescription (bottom panel of
Fig.~\ref{fig:hc_multiRes}), which is much larger than our estimated
error in $f_{\rm peak}$ for either parametrization. Many studies have shown
that $f_{\rm peak}$ is a sensitive probe of the neutron star structure, e.g.
 of the radius or tidal deformability of the initial stars
\cite{Baiotti2017,Paschalidis2017,Bauswein2019,Bernuzzi2020,Radice2020}.
However, as we show in Appendix~\ref{sec:EOSfit}, the mass-radius and
tidal deformability curves for the GPP and PWP parametrizations of the
ENG EoS used in this paper are nearly identical.  Indeed, the
differences in the radius and tidal deformability for a 1.4~$\Ms$ star
between the two parametrizations are only $\Delta R_{1.4} = 0.01$~km
(0.1\%) and $\Delta \Lambda_{1.4} = 4.6$ (1\%), respectively. Thus,
the difference in $f_{\rm peak}$ cannot be easily explained away in terms of
differences in these macroscopic properties.

Instead, this difference may point to a dependence of $h_c$ on the
smoothness of the underlying EoS; for example, through the different
sound speed gradients that must be resolved for each
parametrization. However, it remains possible that this difference in
$f_{\rm peak}$ is simply a result of non-convergence in the
post-merger gravitational waves (as found in Fig.~\ref{fig:conv_phi}).
Additionally, as we discussed in Sec.~\ref{sec:conv}, we find hints
that our lowest resolution evolution is not yet in the convergent
regime for all times. Thus, it is possible that the observed
difference in $f_{\rm peak}$ between the GPP and PWP models may go
away with increased resolution; however, such a study is beyond the
feasibility of the present work.

\subsection{Spurious heating with the PWP and GPP parametrizations}

Finally, we briefly discuss the impact of the EoS parametrization on
the spurious heating of the neutron star surface during the inspiral. Spurious
heating at the neutron star surface is a common feature of merger
simulations, and is caused by spurious shocks that form across the steep
density gradient of the stellar surface. This heating can raise the
pre-merger stellar surface to significant temperatures, but the
heating is typically limited to low-density regions and thus has
little effect on the inspiral dynamics.  In addition, the shock
heating that develops at first-contact during merger is much stronger
than the spurious inspiral heating. As a result, the inspiral
heating is often simply ignored, even though it can hide interesting
physical processes, such as the melting of the neutron star crust
\cite{Hammond2021}.

One might naturally wonder whether using a smooth EoS can reduce this
spurious heating during the inspiral, compared to simulations that
use a PWP EoS with large jumps in the sound speed.  We find that this
is indeed the case.

\begin{figure}[!ht]
\centering
\includegraphics[width=0.475\textwidth]{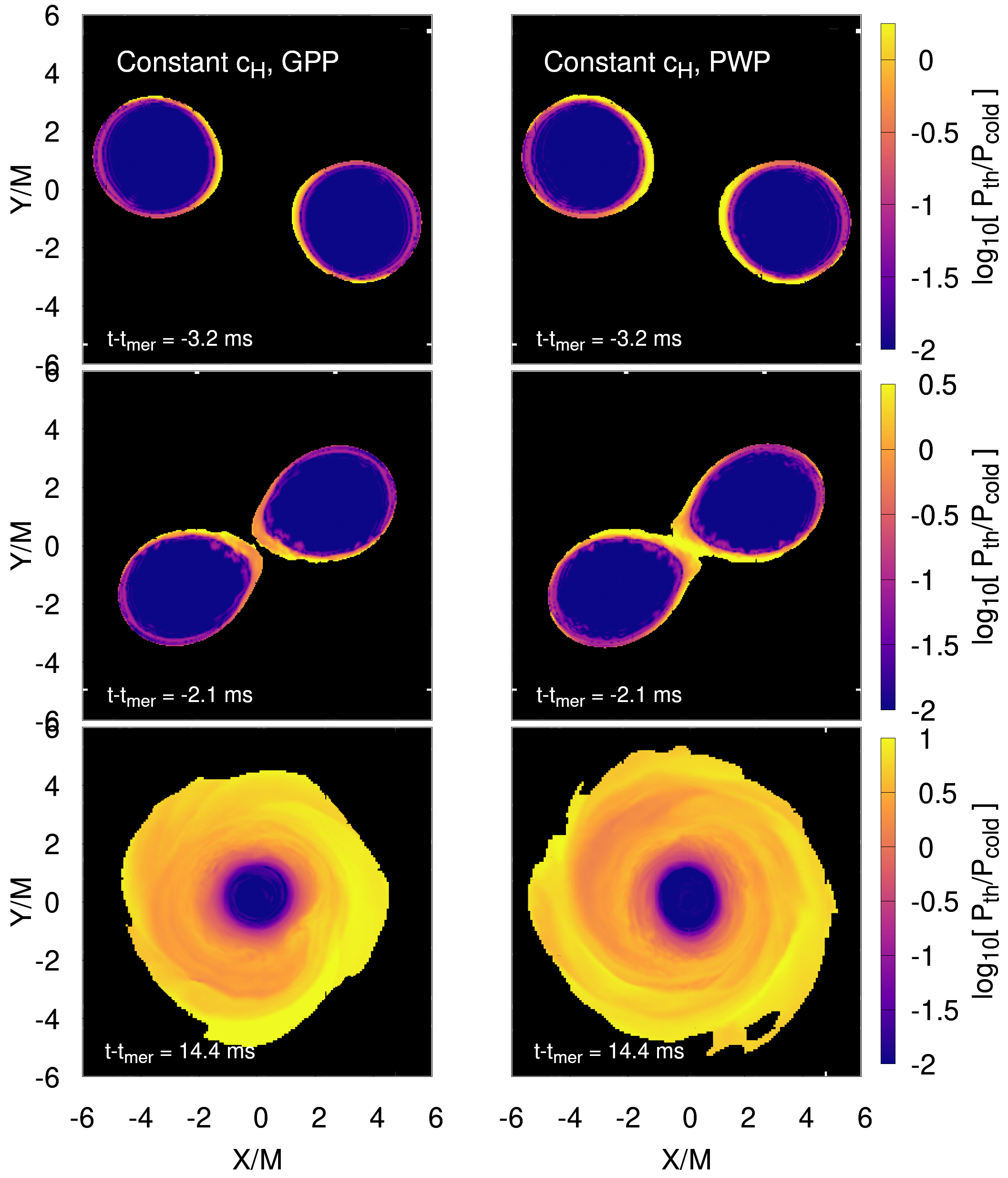}
\caption{\label{fig:Pth_2d} Equatorial snapshots showing the thermal
  pressure relative to the cold pressure, at two times before merger
  and at the end of the simulations. All times are labeled relative to
  the time of merger, $t_{\rm mer}$.  The left column shows the
  results from the evolution with the generalized PWP EoS and the
  right column shows the results with the standard piecewise
  polytropic version of the EoS.  In both cases, the constant $c_H$
  damping prescription is used. Only matter with densities above
  0.01$\times n_{\rm sat}$ (where $n_{\rm sat}=0.16~\text{fm}^{-3}$ is
  the nuclear saturation density) are plotted; the black background is
  included for visual clarity. The PWP parametrization leads to more
  significant spurious heating near the neutron star surface during the
  inspiral. }
\end{figure}

\begin{figure*}[!ht]
\centering
\includegraphics[width=0.9\textwidth]{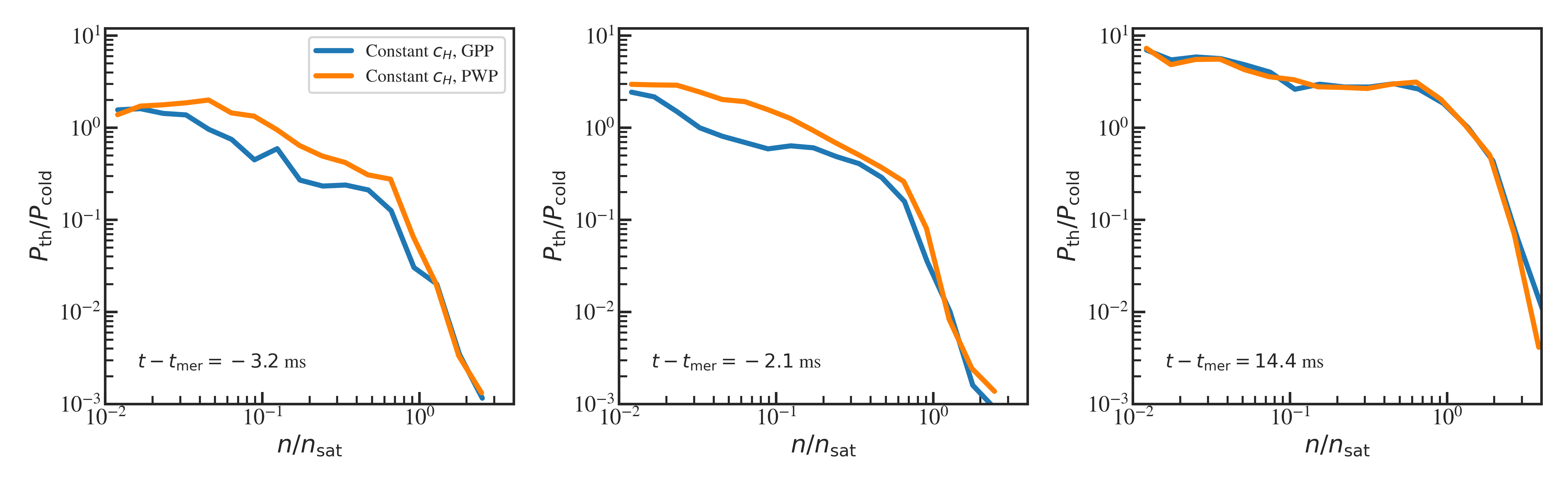}\\
\includegraphics[width=0.9\textwidth]{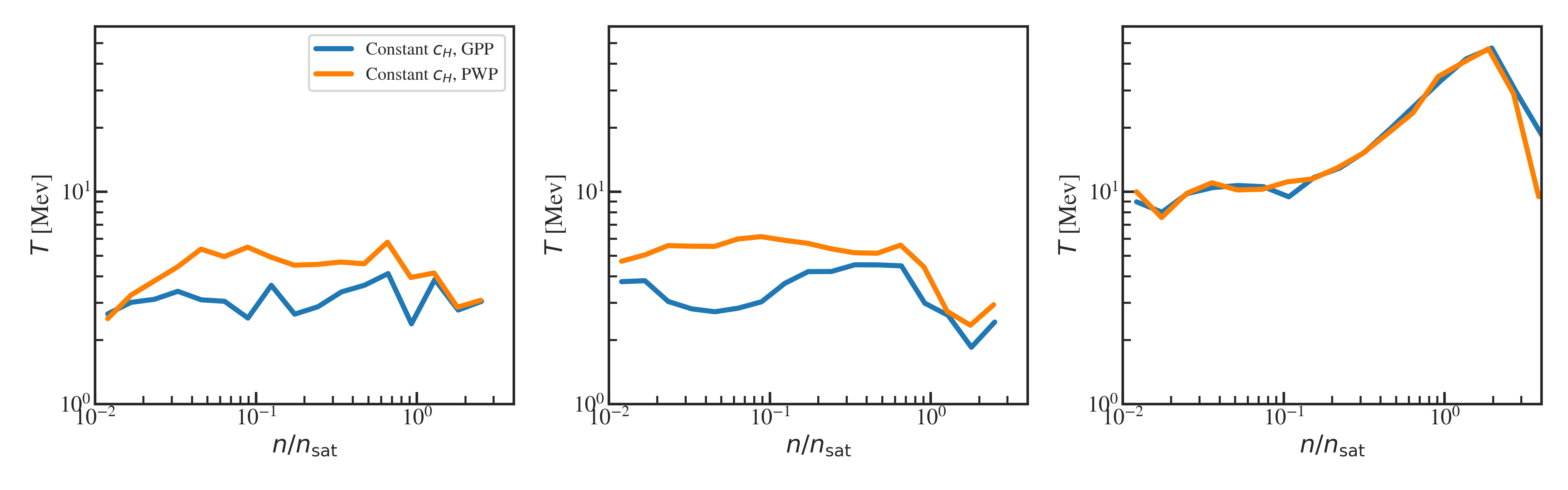}
\caption{\label{fig:T_Pth} Median thermal properties, calculated within density
bins that are log-uniformly spaced, at two times pre-merger and one time post-merger.
These times correspond to the same snapshots as shown in Fig.~\ref{fig:Pth_2d}. The top panel
shows the median value of $P_{\rm th}/P_{\rm cold}$ at each density, while
the bottom panel shows the corresponding, median temperature.
The PWP parametrization
(orange lines) leads to greater heating of the neutron star surface 
during inspiral compared to the GPP parametrization (shown in blue),
where the surface corresponds to densities of $n/n_{\rm sat}\approx0.01-0.5$
(defined as the volume containing 0.99-0.9999 of the total mass for a TOV star
with this EoS).
Both parametrizations give similar thermal profiles for the late time 
remnant (far right column). }
\end{figure*}

Figure~\ref{fig:Pth_2d} shows the 2D thermal profiles
of the neutron stars at two times during the inspiral 
and at one time near the end of our simulations,
for the GPP and PWP EoS parametrizations. In both cases,
the constant $c_H$ damping prescription is used. 
Figure~\ref{fig:Pth_2d}  shows
the thermal pressure, $P_{\rm th}$, relative to the cold pressure, $P_{\rm cold}$,
for all matter with densities above 0.01~$n_{\rm sat}$, where 
$n_{\rm sat}=0.16~\text{fm}^{-3}$ is the nuclear saturation density. Matter
at lower densities (where $P_{\rm cold}$ drops precipitously) is masked in black,
so as not to saturate the color scale. Finally, we note that the colorbar 
varies for the three snapshots, in order to account for
the increased spurious heating throughout the evolution.

In the two pre-merger snapshots, we indeed find differences in the
spurious heating of the neutron star surface, with the piecewise polytropic parametrization
leading to higher values of $P_{\rm th}/P_{\rm cold}$ in the surface
layers of each neutron star. The spurious heating is reduced, though
still present, with the use of the generalized PWP parametrization.
In the post-merger snapshot, the thermal profiles of the remnants are
similar.

In order to explore these spatial profiles in more detail,
Fig.~\ref{fig:T_Pth} shows the median thermal properties of the matter
as a function of density, where the median values are calculated
within density bins that are log-uniformly sampled. The top row of
Fig.~\ref{fig:T_Pth} shows the median values of $P_{\rm th}/P_{\rm
  cold}$, while the bottom row shows the median temperatures.  
 We focus in particular on the surface region in these plots,
 which we define as the range of densities from
  $n/n_{\rm sat} \approx 0.01-0.5$. This range corresponds to 
a volume containing 0.99 to 0.9999 of the total mass for a TOV star 
with the same EoS.  
  During
the inspiral, $P_{\rm th}/P_{\rm cold}$ is approximately twice as
large over the neutron star surface when the piecewise
 polytropes are used, compared to the GPP
parametrization. 
 During the inspiral, the surface temperatures
likewise vary by a factor of $\sim1.5$, with the PWP paramerization
again leading to higher temperatures. In contrast, at late times, the
thermal profiles of the remnant are very similar between the two
parametrizations. In the post-merger remnant, we only find differences
at the very highest densities, where the temperature varies by a
maximum factor of $\sim2$. In this inner core region, the thermal
pressure is subdominant to the cold pressure (i.e., $P_{\rm th} < 0.01
P_{\rm cold}$), so the dynamical impact of this difference is likely
to be small.  However, such differences in the high-density
temperature profiles could be important in determining when
bulk-viscous effects become important 
\cite[e.g.,][]{Alford_bulk_viscosity_PhysRevLett.120.041101,Most2021a}, or in
determining the long-term cooling of the remnant and neutrino
irradiation of the disk.

 \section{Discussion and Conclusions}

In this work, we have investigated two ways of improving the
convergence order in several quantities of a binary neutron star
merger simulation within the BSSN formulation. First, we have studied
how discontinuities in a constraint damping approach between different
refinement levels can spoil the convergence in the post-merger phase.
In particular, we studied three different treatments for the parabolic
term added to the evolution equation for the BSSN variable $\phi$,
which is used to damp the growth of the Hamiltonian constraint. One
might expect that because this term is supposed to be zero in the
continuum limit, it may not matter for the simulations how the
constraint damping parameter in this approach is chosen. We showed
that the standard ``$\Delta$t-scaled" damping ($c_{H} = 0.08 \Delta t$), which
always respects the strict Courant parabolic stability condition as
resolution increases, leads to non-convergence in $||\mathcal{H}||$,
starting shortly after merger.
Scaling the damping coefficient by the
resolution ($c_{H} = 0.08 \left[\Delta x_{\rm HR} / \Delta x_{i}
  \right] \Delta t$) to ensure the same (refinement level dependent) damping for
all three resolutions improves the convergence of $||\mathcal{H}||$
somewhat; but the convergence order still decays with time with this
prescription and is completely lost by the end of our simulations.
We find that using a constant prescription for the
damping ($c_H = 0.0045$~km at all resolutions and at all points on the
grid) enables us to achieve convergence until late times
post-merger. With this continuous damping prescription, we recover the
expected first-order convergence of $||\mathcal{H}||$ with our
numerical scheme for the post-merger phase until the end of our
simulations. 

In other words,
we find that continuity in the parabolic term used to damp
the Hamiltonian constraint in the BSSN formulation is necessary to
ensure convergence in the post-merger phase of our simulations.

In addition to studying how discontinuities in the parabolic damping
of the Hamiltonian constraint affect the convergence of $||\mathcal{H}||$,
 we have also investigated the role
of discontinuities in the sound speed, using two different
parametrizations of the dense-matter EoS. We compared simulation
results for the standard piecewise polytopic parametrization of the
nuclear EoS ENG, as well as a ``generalized" piecewise polytropic
approximation of the same EoS  \cite{OBoyle2020} ,
which ensures smoothness in the EoS.

In general, we find only small differences between the evolutions with
these two parametrizations. Interestingly, we find negligible
differences between the inspiral gravitational waves for these two
parametrizations, in contrast to previous work \cite{Foucart2019}. As
discussed further in Sec.~\ref{sec:gw}, this may be due to the
different initial binary separation in our simulations, the different
nuclear EoS that is being approximated, the different smooth EoS
parametrization that we use, or the fact that~\cite{Foucart2019} adopts
a spectral method for the spacetime evolution instead of the finite-
difference method used here.  If the difference in parametrization turns out
to be the source of the different results, it would be interesting to
understand what feature of the EoS is causing the reduction of the
phase errors, as this would imply that continuity in the sound speed
alone is insufficient.  Further study will be needed to address these
questions.

We find a small difference in the peak frequency of the
post-merger gravitational waves between the PWP and GPP
parametrizations of the EoS, which is larger than
our approximate estimate of the numerical errors for these evolutions.
This finding may point to a (small) dependence of the
post-merger gravitational wave signal on the smoothness of the
microscopic sound speed, which may be important to take into account
when estimating the errors for numerical simulations that use
parametric EoSs. However, we cannot rule out the possibility that this
is simply an artifact of non-convergence in the post-merger
gravitational wave signals. We leave further exploration of this issue
to future work.

Finally, we explored the impact of the EoS parametrization on the
spurious heating of the neutron star surfaces during the inspiral.
Both the GPP and PWP parametrizations have multiple transition
densities at low densities (for details, see the Appendix and
Refs.~\cite{Read2009,OBoyle2020}).  We find that by using an EoS that
is smooth across these transitions -- i.e., an EoS with no artificial
jumps in the sound speed -- the spurious heating during the inspiral
can be reduced. In the post-merger phase, we do not find significant
differences in the remnant thermal profiles, except at the very
highest densities, where the thermal pressure is subdominant to the
cold pressure.

As a final remark, we note that it is unclear how discontinuities at
the refinement level boundaries in Kreiss-Oliger dissipation affect
our results, especially with regards to the gravitational wave
signals, since these are extracted at large distances from the center
of mass and hence travel across multiple refinement levels to reach
the extraction point. In~\cite{Bozzola2021}, it was demonstrated that
ensuring continuity of the Kreiss-Oliger dissipation operator was
crucial for the stability of simulations of charged black hole
binaries. Such discontinuities could affect other formulations of the
Einstein equations as well, such as the families of Z4 or generalized
harmonic formulations. By making the Kreiss-Oliger dissipation
continuous, it may be possible to further improve the post-merger
convergence in our simulations. This will be the subject of a future
investigation of ours.

\begin{acknowledgments} We are grateful to Michael O'Boyle
for providing us with a high-accuracy table of fitting coefficients
for the low-density GPP representation of SLy and for his correspondence
on fitting for GPP coefficients. We thank William East
and Frans Pretorius for providing us with their generic primitives
recovery routine, and Zach Etienne for useful conversation on this work. 
CR would like to thank Elias Most for insightful
discussions related to this work.  CR gratefully acknowledges
support from a joint postdoctoral fellowship at the Princeton Center
for Theoretical Science, the Princeton Gravity Initiative, and as a
John N. Bahcall Fellow at the Institute for Advanced Study.  This work
was in part supported by NSF Grant PHY-1912619 and PHY-2145421 to the
University of Arizona.  The simulations presented in this work were
carried out in part with the Princeton Research Computing resources at
Princeton University, which is a consortium of groups led by the
Princeton Institute for Computational Science and Engineering
(PIC-SciE) and Office of Information Technology's Research
Computing. The simulations were additionally made possible thanks to
the {\tt Stampede2} cluster at the Texas Advanced Computing Center,
under XSEDE allocation PHY190020.
\end{acknowledgments}

\FloatBarrier
\appendix
\section{Comparison of GPP and PWP EoS approximations}
\label{sec:EOSfit}

  \begin{table*}
  \centering
\begin{tabular}{cccccc || ccc|ccc }
\hline 
EoS  &  $\rho_0$ ($\times 10^{14}$ g/cm$^3$)  &  $\log_{10} K_1$  &   $\Gamma_1$   &  $\Gamma_2$ &  $\Gamma_3$  &    
			$R_{1.4}$           &  \%$_{\text{PWP}}$    &   \%$_{\text{full}}$  & 
		         $\Lambda_{1.4}$ &  \%$_{\text{PWP}}$    &   \%$_{\text{full}}$  
\\
\hline \hline 
ENG & 1.064 &  -34.7162 &   3.277 &  2.863 & 3.272 & 
			11.95 & 0.09 & 1.5  &
			362.5 & 1.2 & 11.3 \\ 
\hline
\end{tabular}
\caption{\label{table:GPP} Generalized piecewise polytrope fit parameters for the ENG EoS. 
$R_{1.4}$ and $\Lambda_{1.4}$ indicate the radius and tidal deformability
of a 1.4~$\Ms$ neutron star, predicted by the GPP EoS. The columns $\%_{\text{PWP}}$ report the percent differences
between the predictions for each of these values, computed by the GPP and the PWP approximations. The column
$\%_{\text{ full}}$ reports the percent differences between the predictions of the GPP approximation and the full EoS.}
 \end{table*}

 \begin{figure*}[!ht]
\centering
\includegraphics[width=0.8\textwidth]{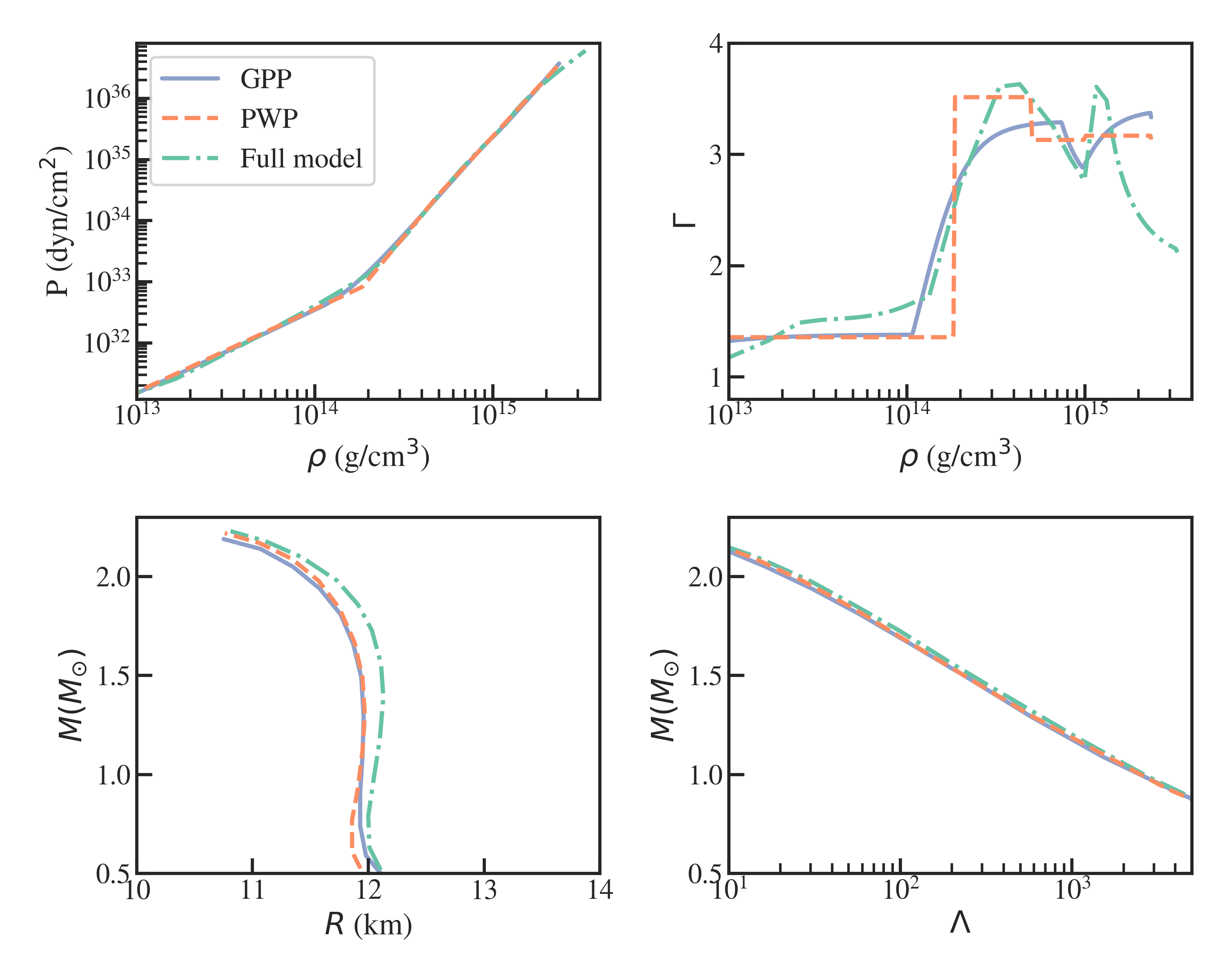}
\caption{\label{fig:eos} Comparison of properties for the full ENG EoS model, the generalized 
PWP approximation (GPP), and the PWP approximation. Clockwise from top left:  pressure $P$ as
a function of density $\rho$; the polytropic index, defined as $\Gamma = \partial \ln P / \partial \ln \rho$;
the tidal deformability as a function of mass; and the mass-radius relation.
 The GPP and PWP give nearly indistinguishable results
for the global properties of the neutron star, with small deviations from the full ENG model due to the different
crust descriptions. }
\end{figure*}

In this appendix, we describe our numerical procedure for creating a generalized
piecewise polytrope (GPP) parametrization of the ENG EoS. 

For the low-density portion of the EoS, we use the GPP representation of SLy,
the coefficients for which are provided in Table II of \cite{OBoyle2020}.\footnote{For this
  work, we use coefficients from a higher-accuracy version of Table II
  which was graciously provided by M. O'Boyle (priv. comm.).} SLy is
used to describe the crust of the EoS, up to the density
at which the crust and high-density EoSs intersect.

For the high-density EoS, we follow \cite{OBoyle2020} in using three piecewise polytropic
segments, which are divided at fiducial densities $\rho_1 = 10^{14.87}$~g/cm$^3$ and 
$\rho_2 = 10^{14.99}$~g/cm$^3$. This leaves us with four free variables: $\{ \overline{K}_1, 
\overline{\Gamma}_1, \overline{\Gamma}_2, \overline{\Gamma}_3 \}$. From these quantities
and the crust coefficients, all other $\overline{K}_i$, $\Lambda_i$ in eqs.~(\ref{eq:P_GPP}-\ref{eq:lambda_GPP}) 
are uniquely determined.

We perform a Markov Chain Monte Carlo simulation to find the set of $\{ \overline{K}_1, 
\overline{\Gamma}_1, \overline{\Gamma}_2, \overline{\Gamma}_3 \}$ that minimizes
the differences between the GPP pressure and the
pressure predicted by the tabulated EoS. We bound the fit
between densities of $10^{14}$~g/cm$^3$ and the density corresponding
to the maximum mass. We calculate an initial guess for the 
parameters by performing a piecewise power-law fit to
$\partial P/\partial \rho$, and iterate from these values
for a total of 50,000 iterations. The resulting best-fit parameters are
reported in Table~\ref{table:GPP}, while the agreement
between the tabulated ENG pressure, the standard PWP approximation,
and our new GPP fit is shown in the top left panel of Fig.~\ref{fig:eos}. 
We find only very small differences in the pressures predicted by
either approximation of the EoS and the full model.

The top right panel of Fig.~\ref{fig:eos} shows the adiabatic index for
each of these representations of ENG. The adiabatic index of the complete
model is smooth, as expected for this nuclear EoS which does not
undergo any physical phase transition at high densities. The PWP
approximation, in contrast, shows large discontinuities at each of the
fiducial densities dividing the polytropic segments. The adiabatic
index with the GPP approximation is again smooth, as intended.

Finally, we also calculate the mass-radius and mass-tidal deformability curves 
for each representation of the EoS, which we show in the bottom row
of Fig.~\ref{fig:eos}. The PWP and GPP representations are nearly identical to 
one another, and show only a small offset from the full model. This offset
stems from the differences in the crust EoS:  both approximations
switch to the SLy EoS at low densities, whereas the ``full model"
in Fig.~\ref{fig:eos} corresponds to ENG at all densities. 

We also report the radius and
tidal deformability of a 1.4~$\Ms$ neutron star predicted the GPP approximation
with our best-fit parameters in Table~\ref{table:GPP}. For the purposes of this paper, 
the most relevant comparison is that between the PWP and GPP approximations,
but we also report the differences between the GPP approximation and full EoS for completeness. 
We find a fractional difference in $R_{1.4}$ of only 0.09\% between the two parametrizations,
and a fractional difference of $1.2\%$ in $\Lambda_{1.4}$. Thus, 
the global properties predicted by the GPP and PWP approximations
are nearly indistinguishable, which allows us to directly study
the impact of the sound speed treatment on our numerical simulations,
without the confounding variable of changes to the macroscopic stellar structure.

\FloatBarrier

\bibliography{gwthermal}
\bibliographystyle{apsrev4-1}

\end{document}